\documentclass[a4paper,11pt]{article}
\usepackage{ifpdf}

\usepackage{epsfig,multicol,amsmath}
\usepackage[T1]{fontenc}

\usepackage{jheppub}
\usepackage{epstopdf}

\usepackage{bm}  
\usepackage{amsmath}     
\usepackage{epsfig,epsf}
\usepackage{graphicx}
\usepackage{rotating}
\usepackage{verbatim}

\def\beq{\begin{equation}}
\def\eeq{\end{equation}}
\def\bea{\begin{eqnarray}}
\def\eea{\end{eqnarray}}

\def\eq#1{{Eq.~(\ref{#1})}}
\def\fig#1{{Fig.~\ref{#1}}}

\newcommand{\Lb}{\left(}
\newcommand{\Rb}{\right)}
\parskip=\itemsep               

\newcommand{\nn}{\nonumber}

\newcommand{\h}{\frac{1}{2}}

\newcommand{\vstrut}{\rule[-7pt]{0pt}{1pt}}
\newcommand{\pom}{I\!\!P}

\def\pom{{I\!\!P}}
\def\reg{{I\!\!R}}
\title{ CGC/saturation approach for soft interactions at high energy:
 long range rapidity correlations}
\author[a]{E. ~Gotsman,}
\author[a,b]{ E.~ Levin}
\author[a]{  and U.~ Maor}

\affiliation[a]{Department of Particle Physics, School of Physics and Astronomy,
Raymond and Beverly Sackler
 Faculty of Exact Science, Tel Aviv University, Tel Aviv, 69978, Israel}
\affiliation[b]{Departemento de F\'isica, Universidad T\'ecnica Federico Santa Mar\'ia, and Centro Cient\'ifico-\\
Tecnol\'ogico de Valpara\'iso, Avda. Espana 1680, Casilla 110-V, Valpara\'iso, Chile}
\emailAdd{gotsman@post.tau.ac.il}
\emailAdd{leving@post.tau.ac.il, eugeny.levin@usm.cl}
\emailAdd{maor@post.tau.ac.il}
\abstract{ In this paper we continue our program to build a model for
 high energy soft interactions, that is based on the CGC/saturation 
approach.
The main result of this paper is that we have discovered a mechanism that 
leads to
  large long range rapidity correlations, and results in 
  large values of the correlation function
 $R\Lb y_1,y_2\Rb \,\geq \,1$,  which is
  independent of $y_1$ and $ y_2$. Such behaviour of the  correlation 
function, 
provides 
  strong support for the idea, that at high energies
 the system of partons that is produced, is not
  only dense, but also has  strong attractive forces acting between the 
partons.
}

\keywords{BFKL Pomeron, soft interaction, CGC/saturation approach, correlations}

\preprint{TAUP - 3000/15\\
\today}

\begin{document}
\maketitle
\flushbottom

\section{Introduction}
The large body of  experimental data  on soft interactions at
 high 
energy\cite{ALICE,ATLAS,CMS,TOTEM,ALICEI,CMSI,ATLASI,CMSMULT,ATLASCOR},
presently,
 cannot be comprehended  in terms of theoretical 
  high energy QCD (see \cite{KOLEB} for the review).

 In this paper we continue our effort\cite{GLMNI,GLM2CH,GLMINCL} to 
comprehend such interactions, by
  constructing  a model that incorporates  the
 advantages of two theoretical approaches to high energy QCD. 
 
 The first one is the CGC/saturation approach
 \cite{GLR,MUQI,MV,B,MUCD,K,JIMWLK},  which provides a clear picture
 of  multi particle production at high energy, that proceeds in two 
stages.
 The first stage is the production of a mini-jet with the typical 
transverse
 momentum $Q_s$. Where   $Q_s$ the saturation scale,  is much 
larger than the
 soft scale. This stage is under full theoretical control. The second stage is
 when the mini-jet decays into hadrons,
which we have to  treat phenomenologically,  using data from  
hard
 processes. Such an approach leads to a good description of the 
experimental
 data on inclusive production,  both for hadron-hadron, hadron-nucleus and
 nucleus-nucleus collisions, and the observation of some regularities in 
the data,
 such as  geometric scaling behaviour\cite{KLN,KLNLHC,LERE,MCLPR,PRA}. 
 The shortcoming of this approach is, the fact that it is disconnected 
from
  diffractive physics.

On the other hand, there exists a different approach to high energy QCD: 
the
 BFKL Pomeron\cite{BFKL}  and  its 
interactions\cite{GLR,MUPA,BART,BRN,KOLE,LELU,LMP,AKLL,LEPP}, which is
 suitable to describe 
 diffractive physics.   The BFKL Pomeron calculus turns
 out to be  close to the the old  Reggeon theory \cite{COL}, so 
     for
calculating the
 inclusive characteristics of  multiparticle production,
 we can apply the Mueller diagram technique\cite{MUDI}.   
  The relation
 between these two approaches has not yet been established,  but
 they are equivalent\cite{AKLL} for the rapidities ($\ln \Lb s/s_0\Rb$), 
such that 
\beq \label{I1}
Y \,\leq\,\frac{2}{\Delta_{\mbox{\tiny BFKL}}}\,\ln\Lb \frac{1}{\Delta^2_{\mbox
{\tiny BFKL}}}\Rb
\eeq
where $\Delta_{\mbox{\tiny BFKL}}$ denotes the intercept of the BFKL 
Pomeron.
 As we have discussed \cite{GLMNI}, the parameters of our model are
  such that for $Y\, \leq\,36$,
 we can  trust our approach, based on the BFKL 
Pomeron calculus.

 This paper is the next step in our program to build a model for high
 energy soft scattering, based on an analytical calculation, without 
 using a Monte Carlo simulation. We discuss the correlation function:
 \beq \label{R}
  R\Lb y_1, y_2; Y\Rb\,\,=\,\ \frac{1}{\sigma_{NSD}}\frac{d^2 \sigma}{
 d y_1 d y_2}\Bigg{/}\Lb\frac{1}{\sigma_{NSD}}\frac{d \sigma}{ d y_1}
 \,\frac{1}{\sigma_{NSD}}\frac{d \sigma}{ d Y_2}\Rb\,\,-\,\,1
  \eeq
  where $Y$ denotes the total rapidity ($Y = \ln\Lb s/s_0\Rb$ and $s = 
W^2$,
  $W$ is the energy in c.m.f.) and $y_1$ and $y_2$ are the rapidities
 of the produced hadrons. $\sigma_{NSD} \,=\,\sigma_{tot} \,-\,\sigma_{el}
 \,-\,\sigma_{sd}\,-\,\sigma_{dd}$ where $ \sigma_{tot}(\sigma_{el},
\sigma_{sd}, \sigma_{dd} )$  are total, elastic, single and double
 diffraction cross sections.

 The  paper is organized as follows. In the following section we discuss 
the main 
features of our approach, concentrating on the description of
  diffractive processes. In section 3, we derive the main
 formulae for the correlation functions in our approach, while
 in section 4 we compare our predictions with the available experimental 
data.

  \section{Our model: generalities, elastic amplitude and inclusive production}
\begin{figure}[ht]
    \centering
  \leavevmode
      \includegraphics[width=14cm]{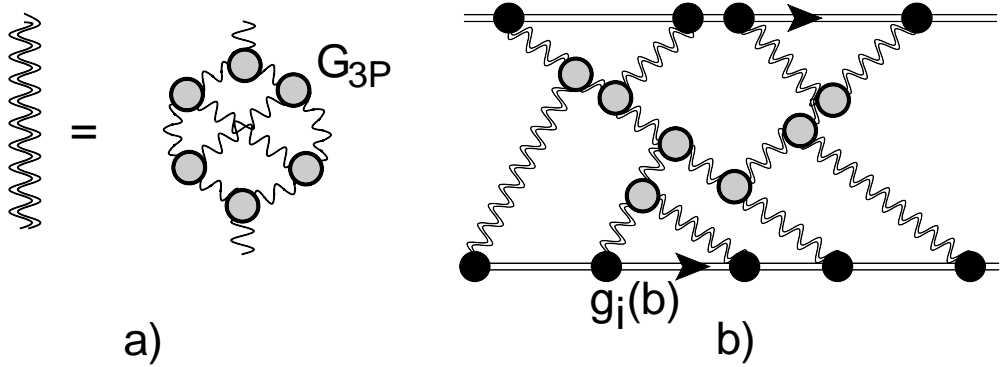}  
      \caption{\fig{amp}-a shows the set of the diagrams in
 BFKL Pomeron calculus. The wavy double lines denote the resulting 
(dressed) Green's
 function of the Pomeron in the framework of high energy QCD,
 while the single wavy lines describe the BFKL Pomerons.
 In \fig{amp}-b we show the net diagrams  that   include the
 interaction of  BFKL Pomerons with colliding hadrons. }
\label{amp}
   \end{figure}

  In this section we briefly review  our model, which 
successfully describes diffractive\cite{GLMNI,GLM2CH} and inclusive
 cross sections\cite{GLMINCL}. The main ingredient of our model is
 the BFKL Pomeron  Green's function, which we determined using the 
CGC/saturation
 approach\cite{GLMNI,LEPP}. We determined this function  from the 
solution of
 the non-linear Balitsky-Kovchegov equation\cite{B,K},  using the MPSI
 approximation\cite{MPSI} to sum enhanced diagrams shown in \fig{amp}-a.
 It has the following form:
 \bea \label{G}
G^{\mbox{\tiny dressed}}\Lb T\Rb\,\,&=&\,\,a^2 (1 - \exp\Lb -T\Rb )  +
 2 a (1 - a)\frac{T}{1 + T} + (1 - a)^2 G\Lb T\Rb \nn\\
~~~&\mbox{with}&~~G\Lb T\Rb = 1 - \frac{1}{T} \exp\Lb \frac{1}{T}\Rb
 \Gamma_0\Lb \frac{1}{T}\Rb
\eea

\beq \label{T}
T\Lb s, b\Rb\,\,=\,\,\phi_0 S\Lb b, m\Rb e^{0.63\lambda \ln(s/s_0)}
~~~~\mbox{with}~~~S\Lb b , m \Rb \,\,=\,\,\frac{m^2}{2 \pi} e^{ - m b}
\eeq 
 In these formulae we take $a=0.65$,  this value was chosen, so as to 
obtain
 the analytical form for the solution of the BK equation. Parameters
 $\lambda$ and $\phi_0$, can be estimated in the leading order of  QCD,
 but due to large next-to-leading order corrections, we treat 
them as parameters of  the fit. $m$ is a non-perturbative parameter,
 which characterize the large impact parameter behavior of the
 saturation momentum, as well as the typical sizes of dipoles that
 take part in the interactions. The value of $m =5.25\,GeV$ in our
 model, justifies our main assumption, that  BFKL Pomeron calculus
 based on a perturbative QCD approach, is able to describe  soft 
physics,
 since $m \,\gg\,\mu_{soft}$, where $\mu_{soft}$ denotes the natural scale 
for
 soft processes ($ \mu_{soft} \,\sim\,\Lambda_{QCD}$ and/or  pion mass).
 
 Unfortunately, since the confinement problem is far from being 
solved, we assume
 a phenomenological approach for the structure of the colliding hadron.
 We use a two channel model, which allows us to calculate the
 diffractive production in the region of small masses.
   In this model, we replace the rich structure of the 
 diffractively produced states, by a single  state with the wave 
function 
$\psi_D$, a la Good-Walker\cite{GW}.
  The observed physical 
hadronic and diffractive states are written in the form 
\beq \label{MF1}
\psi_h\,=\,\alpha\,\Psi_1+\beta\,\Psi_2\,;\,\,\,\,\,\,\,\,\,\,
\psi_D\,=\,-\beta\,\Psi_1+\alpha \,\Psi_2;~~~~~~~~~
\mbox{where}~~~~~~~ \alpha^2+\beta^2\,=\,1;
\eeq 

Functions $\psi_1$ and $\psi_2$  form a  
complete set of orthogonal
functions $\{ \psi_i \}$ which diagonalize the
interaction matrix ${\bf T}$
\beq \label{GT1}
A^{i'k'}_{i,k}=<\psi_i\,\psi_k|\mathbf{T}|\psi_{i'}\,\psi_{k'}>=
A_{i,k}\,\delta_{i,i'}\,\delta_{k,k'}.
\eeq
The unitarity constraints take  the form
\beq \label{UNIT}
2\,\mbox{Im}\,A_{i,k}\left(s,b\right)=|A_{i,k}\left(s,b\right)|^2
+G^{in}_{i,k}(s,b),
\eeq
where $G^{in}_{i,k}$ denotes the contribution of all non 
diffractive inelastic processes,
i.e. it is the summed probability for these final states to be
produced in the scattering of a state $i$ off a state $k$. In \eq{UNIT} 
$\sqrt{s}=W$ denotes the energy of the colliding hadrons, and $b$ 
the 
impact  parameter.
A simple solution to \eq{UNIT} at high energies, has the eikonal form 
with an arbitrary opacity $\Omega_{ik}$, where the real 
part of the amplitude is much smaller than the imaginary part.
\beq \label{A}
A_{i,k}(s,b)=i \Lb 1 -\exp\Lb - \Omega_{i,k}(s,b)\Rb\Rb,
\eeq
\beq \label{GIN}
G^{in}_{i,k}(s,b)=1-\exp\Lb - 2\,\Omega_{i,k}(s,b)\Rb.
\eeq
\eq{GIN} implies that $P^S_{i,k}=\exp \Lb - 2\,\Omega_{i,k}(s,b) \Rb$, is 
the probability that the initial projectiles
$(i,k)$  reach the final state interaction unchanged, regardless of 
the initial state re-scatterings.
\par

Note, that there is no factor $1/2$, its absence stems from our definition
 of the dressed Pomeron.


\begin{table}[h]
\begin{tabular}{|l|l|l|l|l|l|l|l|l|l|}
\hline
model &$\lambda $ & $\phi_0$ ($GeV^{-2}$) &$g_1$ ($GeV^{-1}$)&$g_2$ ($GeV^{-1}$)& $m(GeV)$ &$m_1(GeV)$& $m_2(GeV)$ & $\beta$& $a_{\pom \pom}$\\
\hline
 2 channel & 0.38& 0.0019 & 110.2&  11.2 & 5.25&0.92& 1.9 & 0.58 &0.21 \\
\hline
\end{tabular}
\caption{Fitted parameters of the model. The values are taken 
from Ref.\cite{GLM2CH}.}
\label{t1}
\end{table}
In the eikonal approximation we replace $ \Omega_{i,k}(s,b)$ by 
\beq \label{EAPR}
 \Omega_{i,k}(s,b)\,\,=\,\int d^2 b'\,d^2 b''\, g_i\Lb \vec{b}'\Rb \,G^{\mbox{\tiny dressed}}\Lb T\Lb s, \vec{b}''\Rb\Rb\,g_k\Lb \vec{b} - \vec{b}'\ - \vec{b}''\Rb 
 \eeq
 We propose a more general approach, which takes into account new
 small parameters, that come from the fit to the experimental data
 (see Table 1 and \fig{amp}):
 \beq \label{NEWSP}
 G_{3\pom}\Big{/} g_i(b = 0 )\,\ll\,\,1;~~~~~~~~ m\,\gg\, m_1 
~\mbox{and}~m_2
 \eeq
 
 The second equation in \eq{NEWSP} leads to the fact that $b''$ in 
\eq{EAPR} is much
 smaller that $b$ and $ b'$  therefore, \eq{EAPR} can be re-written in
 a simpler form
 \bea \label{EAPR1}
 \Omega_{i,k}(s,b)\,\,&=&\,\Bigg(\int d^2 b''\,G^{\mbox{\tiny dressed}}\Lb
 T\Lb s, \vec{b}''\Rb\Rb\Bigg)\,\int d^2 b' g_i\Lb \vec{b}'\Rb \,g_k\Lb
 \vec{b} - \vec{b}'\Rb \,\nn\\
 &=&\,\tilde{G}^{\mbox{\tiny dressed}}\Lb\bar{T}\Rb\,\,\int d^2 b' g_i\Lb
 \vec{b}'\Rb \,g_k\Lb \vec{b} - \vec{b}'\Rb \eea
 
 Selecting the diagrams using the first equation in \eq{NEWSP}, indicates
 that the main contribution stems from the net diagrams shown in \fig{amp}-b.
 The sum of these diagrams\cite{GLM2CH} leads to the following expression 
for $
 \Omega_{i,k}(s,b)$
 \bea
\Omega\Lb Y; b\Rb~~&=& ~~ \int d^2 b'\,
\,\,\,\frac{ g_i\Lb\vec {b}'\Rb\,g_k\Lb\vec{b} -
 \vec{b}'\Rb\,\tilde{G}^{\mbox{\tiny dressed}}\Lb T\Rb
}
{1\,+\,G_{3\pom}\,\tilde{G}^{\mbox{\tiny dressed}}\Lb T\Rb\left[
g_i\Lb\vec{b}'\Rb + g_k\Lb\vec{b} - \vec{b}'\Rb\right]} ;\label{OM}\\
g_i\Lb b \Rb~~&=&~~g_i \,S_p\Lb b; m_i \Rb ;\label{g}
\eea
where
$$
S_p\Lb b,m_i\Rb\,=\,\frac{1}{4 \pi} m^3_i \,b \,K_1\Lb m_i b \Rb
$$
$$
\tilde{G}^{\mbox{\tiny dressed}}\Lb \bar{T}\Rb\,\,=\,\,\int d^2 b
 \,\,G^{\mbox{\tiny dressed}}\Lb T\Lb s, b\Rb\Rb
$$
where $ T\Lb s, b\Rb$ is given by \eq{T}.

Note  that  $\bar{G}^{\mbox{\tiny dressed}}\Lb \bar T\Rb$ does not depend
 on $b$ and is a function of $
\bar{T} =T\Lb s, b=0 \Rb\,=\,\phi_0 \,e^{0.63\,\lambda Y}$.

In all previous formulae, the value of the triple BFKL Pomeron vertex
 is known: $G_{3 \pom} = 1.29\,GeV^{-1}$.

\begin{figure}[ht]
    \centering
  \leavevmode
      \includegraphics[width=6cm]{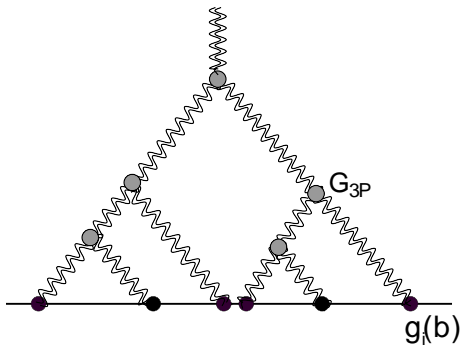}  
      \caption{A typical example of  `fan' diagrams that
 are summed in \protect\eq{NBK}.  }
\label{bk}
   \end{figure}

To simplify further discussion, we introduce the notation 

 \beq \label{NBK}
N^{BK}\Lb G^i_\pom\Lb Y,b \Rb\Rb \,\,=\,\,a\,\Lb 1
 - \exp\Lb -  G^i_\pom\Lb Y, b\Rb\Rb\Rb\,\,+\,\,\Lb 1 - a\Rb
\frac{ G^i_\pom\Lb  Y, b\Rb}{1\,+\, G^i_\pom\Lb Y, b\Rb},
\eeq 
 with $a = 0.65$ .
 \eq{NBK} is an analytical approximation to the numerical solution for  
the 
BK equation\cite{LEPP}. $G_\pom\Lb Y; b\Rb \,=\,\,
 g_i\Lb b \Rb \,\tilde{G}^{\mbox{\tiny dressed}}\Lb \bar{T}\Rb $.
 We recall that the BK equation sums the `fan'  diagrams shown in 
\fig{bk}.

For the  elastic amplitude we have

\beq \label{EL}
a_{el}(b)\,=\,\Lb \alpha^4 A_{1,1}\,
+\,2 \alpha^2\,\beta^2\,A_{1,2}\,+\,\beta^4 A_{2,2}\Rb. 
\eeq

\begin{figure}[ht]
    \centering
  \leavevmode
      \includegraphics[width=5cm]{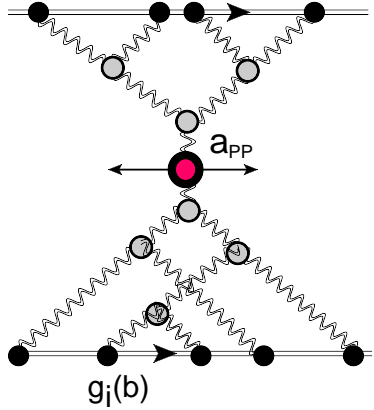}  
      \caption{ Mueller diagram for the  single inclusive cross section.
  The  double wavy lines describe  the resulting Green's function of
 the BFKL Pomerons ( $\tilde{G}^{\mbox{\tiny dressed}} $). The blobs
 stand for the vertices which are  the same as in \fig{amp}.}
\label{incl}
   \end{figure}

To determine  the correlation function (given  in \eq{R}),  we need 
to know
 the 
single inclusive cross sections.  We have discussed these cross sections
 in Ref.\cite{GLMINCL},  for the sake of completeness we give the formula
 that describes the Mueller diagram of \fig{incl}.

 \bea \label{INCF}
   \frac{d \sigma}{d y}\,\,&=&\,\,\int d^2 p_T\,\frac{d \sigma}{d y \,d^2 p_T}\,\,=\,\,a_{\pom \pom}\,\ln\Lb W/W_0\Rb\Bigg\{ \alpha^4  \,In^{(1)}\Lb \h Y + y\Rb \, In^{(1)}\Lb \h Y - y\Rb  \,\nn\\
   &+&\,\alpha^2\beta^2 \Big(In^{(1)}\Lb \h Y + y\Rb \, In^{(2)}\Lb \h Y - y\Rb \,+\,  In^{(2)}\Lb \h Y + y\Rb \, In^{(1)}\Lb \h Y - y\Rb\Big)\,\nn\\
   &+&\,\beta^4 \,In^{(2)}\Lb \h Y + y\Rb \, In^{(2)}\Lb \h Y - y\Rb  \Bigg\}
\eea
where $Y$ denotes the total rapidity of the colliding particles,  and $y$
 is the rapidity of produced hadron. $In^{(i)}$ is given by
\beq \label{IN}
In^{(i)}\Lb y\Rb\,=\,\int d^2 b \,\,N^{BK}\Lb g^{(i)}\,S\Lb m_i,
 b\Rb \,\tilde{G}_\pom\Lb y\Rb\Rb
\eeq

$a_{\pom\pom}$ is a fitted parameter, that was determined
 in Ref.\cite{GLMINCL} (see Table 1).
\section{Two particle correlations}

\subsection{Correlations between two parton showers}
\begin{figure}[ht]
    \centering
  \leavevmode
      \includegraphics[width=15cm]{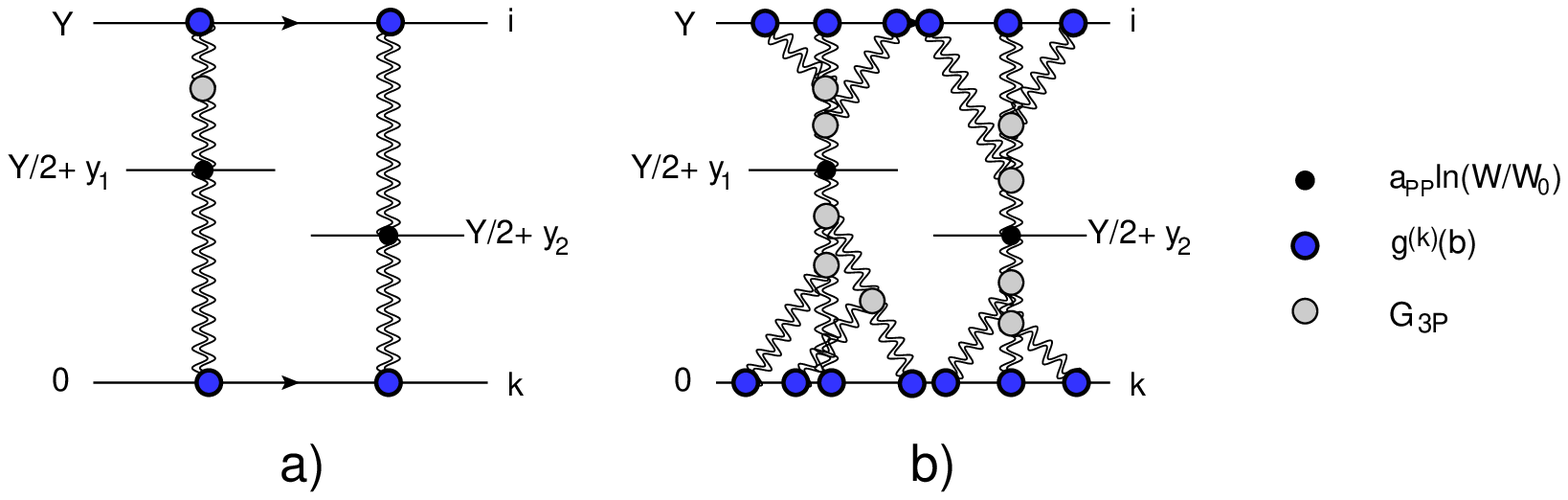}  
      \caption{The  Mueller diagram for the  rapidity correlation between
 two particles produced in two parton showers. \fig{nimcor2sh}-a  shows
 the first Mueller diagram, while \fig{nimcor2sh}-b indicates the 
structure
 of general diagrams.
  The double  wavy lines describe the dressed  BFKL Pomerons.
 The blobs stand for the vertices as shown in the legend.}
\label{nimcor2sh}
   \end{figure}

The Mueller diagram for the correlations between two parton showers is shown
 in \fig{nimcor2sh}. 
Examining this diagram, we see that the contribution to the double
 inclusive cross section, differs from the product of two single inclusive
 cross sections. There are two reasons for this, the first, is that in the 
expression for the double
 inclusive cross section, we integrate  the product of the single
 inclusive inclusive cross sections, over $b$, at fixed $b$. The second, 
is that the 
summation 
 over $i$ and $k$  for  the product of single  inclusive cross
 sections, is for fixed $i$ and $k$.

Introducing the  following new function, enables us to write the 
 analytical expression: 
\bea \label{2SH1}
&&I^{(i,k}\Lb y, b \Rb\,\,=\,\,a_{\pom \pom}\,\ln\Lb W/W_0\Rb\\
&& \times\int d^2 b' \,\,N^{BK}\Bigg( g^{(i)}\,S\Lb m_i,b'\Rb
 \tilde{G}^{\mbox{\tiny dressed}}\Lb\h Y + y
 \Rb\Bigg)\,\,N^{BK}\Bigg( g^{(k)}\,S\Lb m_k,
 \vec{b} - \vec{b}^{\,'}\Rb \tilde{G}^{\mbox{\tiny dressed}}\Lb\h Y- y \Rb\Big)\nn
\eea

 Using \eq{2SH1} we can write the double inclusive cross section in the form
 \bea \label{2SH2}
 &&\frac{d^2 \sigma^{\mbox{\tiny 2 parton showers}}}{ d
 y_1\,\,d y_2}\,\,=\,\,\int d^2 b \,\,\Bigg\{ \alpha^4
 \,I^{(1,1)}\Lb y_1, b \Rb\,I^{(1,1)}\Lb y_2, b \Rb\\
 && \alpha^2\,\beta^2\,\Lb  I^{(1,2)}\Lb y_1, b \Rb
\,I^{(1,2)}\Lb y_2, b \Rb\,\,+\,\,I^{(2,1)}\Lb y_1,
 b \Rb\,I^{(2,1)}\Lb y_2, b \Rb \Rb\,\,+\,\,\beta^4\,I^{(2,2)}\Lb
 y_1, b \Rb\,I^{(2,2)}\Lb y_2, b \Rb \Bigg\}\nn
 \eea
 
 Comparing \eq{2SH2} with   \eq{INCF} squared, we note the
 different powers of $\alpha$ and $\beta$, which reflect the different
 summation over $i$ and $k$, as well as different integration over $b$.
 
 \subsection{Correlations in one parton shower: semi-enhanced diagrams}
 The main theoretical assumption that we make in calculating the 
correlation
 in  a one parton shower, is that  the Mueller diagram technique
 \cite{MUDI}, and the AGK cutting rules\cite{AGK} are valid.  We 
should note, however,
 that even if the Mueller diagrams  provide the correct description of 
 inclusive
 processes in QCD, the AGK cutting rules are not valid   for
 calculations of the correlations in QCD \cite{KOJA,LEPR}. Nevertheless,
 we believe 
  that, we can neglect the AGK cutting rules violating
 contributions since, first, they do not lead to  long range
 rapidity correlations, which are the main subject of our concern,
 and second, as we will show below, the correlations in one parton
 shower turn out to be negligibly small.

 It is instructive to write the expression for the first Mueller diagram 
in
 the following form ( see
 \bea \label{1MDsen}
 &&\frac{d^2 \sigma^{\mbox{one parton shower}}_{senh}}{d y_1\,d y_2}\,\,=\,\,\Lb a_{\pom \pom}\,\ln\Lb W/W_0\Rb\Rb^2\sum_{i=1,k=1}^2\alpha^2_i\,\alpha^2_k\,\int^Y_{\max(\h Y - y_1, \h Y - y_2)}\!\!\!\!\! d Y'\\
 &&\,\int d^2 b \,g_i\Lb b \Rb
 \tilde{G}\Lb Y - Y'\Rb\,G_{3 \pom}\,\tilde{G}\Lb Y' - \h Y - y_1\Rb \,\tilde{G}\Lb Y' - \h Y - y_2\Rb \nn\\
 &&\int d^2 b'  N^{BK}\Bigg(g^k\Lb b \Rb \tilde{G}\Lb \h Y - y_1\Rb\Bigg)\,\int d^2 b'  N^{BK}\Bigg(g^k\Lb b \Rb \tilde{G}\Lb \h Y - y_2\Rb\Bigg)\nn \eea

  The expression for the first Mueller diagram for
 two parton showers correlation (see \fig{nimcor2sh}-a)  has the form
  \bea \label{1MD2sh}
 &&\frac{d^2 \sigma^{\mbox{one parton showers}}_{senh}}{d y_1\,d y_2 }\,\,=\,\  \int d^2 b 
 \frac{d^2 \sigma^{\mbox{2 parton showers}}}{d y_1\,d y_2\,d^2 b} =\,\\
 &&\Lb a_{\pom \pom}\,\ln\Lb W/W_0\Rb\Rb^2\sum_{i=1,k=1}^2\alpha^2_i\,\alpha^2_k\,\int d^2 b \,\Lb g_i\Lb b \Rb\Rb^2
 \tilde{G}\Lb Y' - \h Y - y_1\Rb \,\tilde{G}\Lb Y' - \h Y - y_2\Rb \nn\\
 &&\int d^2 b'  N^{BK}\Bigg(g^k\Lb b \Rb \tilde{G}\Lb \h Y - y_1\Rb\Bigg)\,\int d^2 b'  N^{BK}\Bigg(g^k\Lb b \Rb \tilde{G}\Lb \h Y - y_2\Rb\Big)\nn \eea 
 
\begin{figure}[h]
\centering
\begin{tabular}{c c }
\includegraphics[width=0.4\textwidth, height=3.5cm]{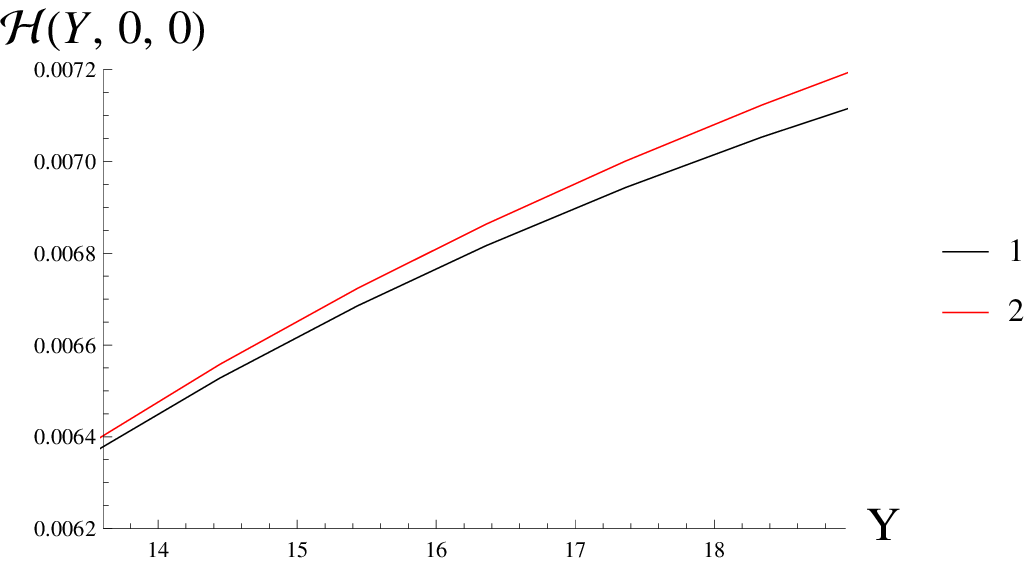}&\includegraphics[width=0.4\textwidth, height=3.5cm]{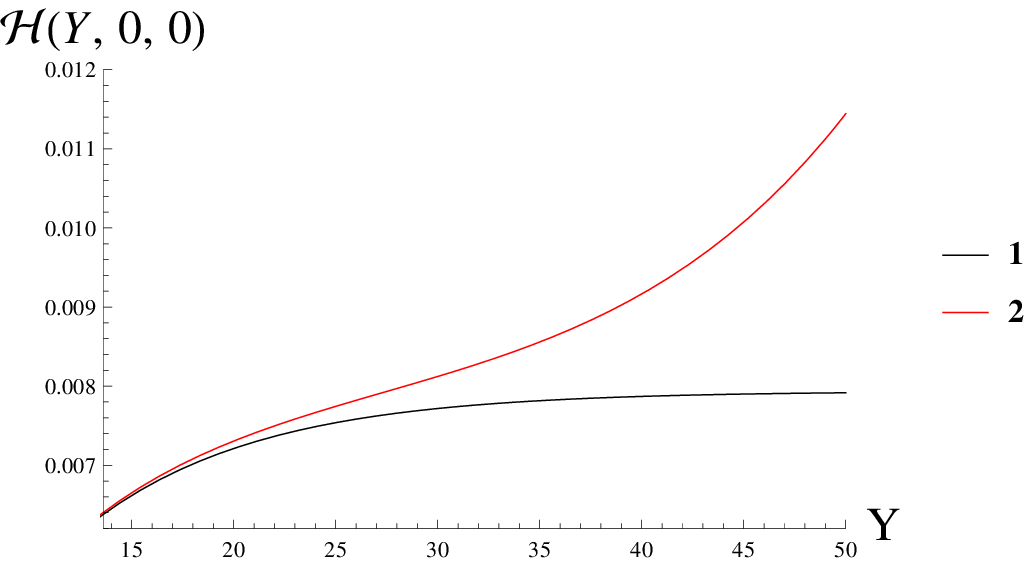}\\
\fig{h}-a &  \fig{h}-b \\
\includegraphics[width=0.4\textwidth,height=3.5cm]{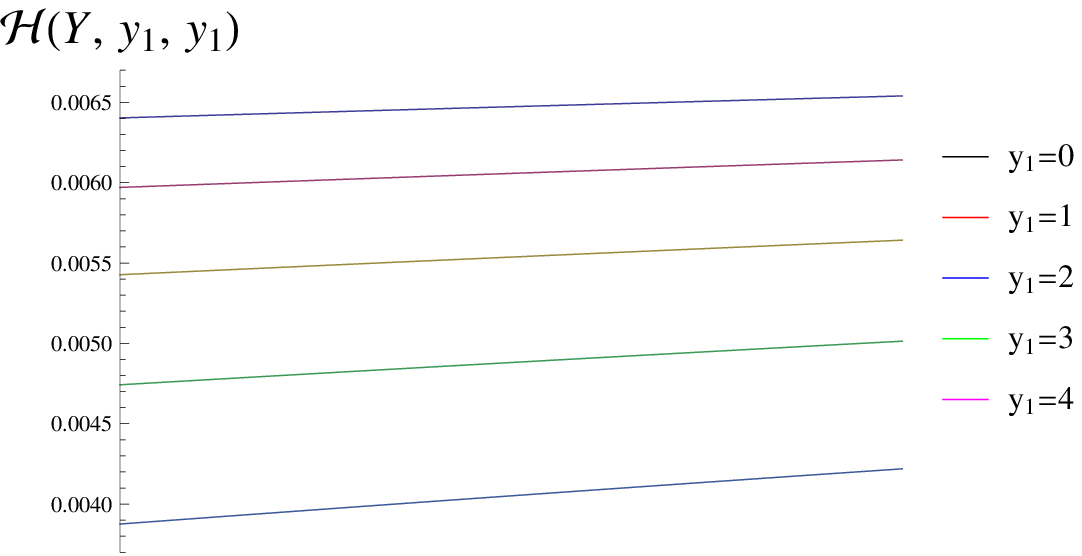}&\includegraphics[width=0.4\textwidth,height=3.5cm]{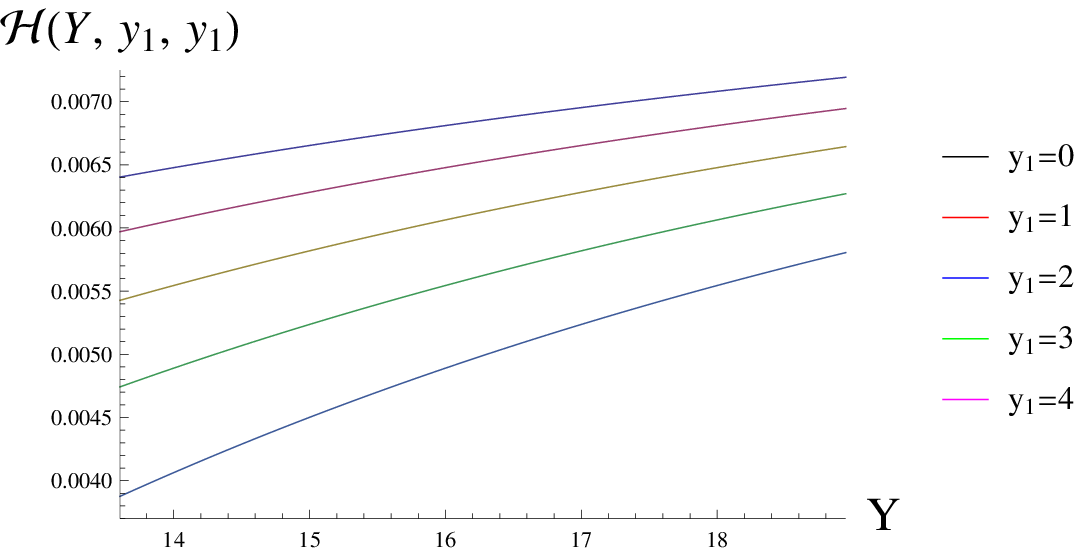}\\
\fig{h}-c &  \fig{h}-d \\
\end{tabular}
\caption{\fig{h}-a: The function ${\cal H}\Lb Y,y_1,y_2\Rb$ versus $Y = 
\ln 
\Lb W^2/W^2_0\Rb$ for $y_1 = y_2=0$.
 The curve 1 shows ${\cal H}\Lb Y,y_1,y_2\Rb$ in
 which all $\tilde{G}$ are replaced by the first
 order of its expansion with respect to $T\Lb Y,
 b\Rb$; line 2 is the exact calculation. \fig{h}-b
 is the same as \fig{h}-a but for a large region of
 $Y$. In \fig{h}-c ${\cal H}\Lb Y,y_1,y_1\Rb$ is plotted
  versus $Y$ at different values of $y_1$. ${\cal H}\Lb
 Y,0,y_2\Rb$ is shown in \fig{h}-d as function of $|y_2|$
 at different energies $W$.}
\label{h}
\end{figure}

 Comparing \eq{1MDsen} with \eq {1MD2sh} one can see that 
 \bea \label{REL12}
 \frac{d^2 \sigma^{\mbox{one parton shower}}_{senh}}{d y_1\,d y_2}\,\,&=&\,\,
 {\cal H}\Lb Y,y_1,y_2\Rb\int d^2 b \,\frac{G_{3 \pom}}
 {g_1\Lb b\Rb} \frac{d^2 \sigma^{\mbox{2 parton showers}}}{d
 y_1\,d y_2\,d^2 b} \\
  {\cal H}\Lb Y,y_1,y_2\Rb\,\,&=&\,\,\int^Y_{\max(\h Y - y_1,
 \h Y - y_2)}\!\!\!\!\! d Y' \frac{ \tilde{G}\Lb Y -
 Y'\Rb\,\tilde{G}\Lb Y' - \h Y - y_1\Rb \,\tilde{G}\Lb
 Y' - \h Y - y_2\Rb}{\tilde{G}\Lb  \h Y - y_1\Rb \,
\tilde{G}\Lb \h Y - y_2\Rb}  \nn
 \eea
 $  {\cal H}\Lb Y,y_1,y_2\Rb$  is proportional to $\h Y - y_1$($y_1 >
 y_2$) in the kinematic region where $\tilde{G}\Lb Y\Rb$ is a constant.
 At small $Y$ it is a constant and is equal to $\int d^2 b T\Lb Y=0, b
 \Rb/\Delta$, where $\Delta = 0.63\,\lambda$. Therefore, we can expect 
that
 the semi-enhanced diagrams can give  larger contribution to the double
 inclusive cross section than the production from two parton showers. 
 However, \fig{h} shows that both the value, and the increase  turns out
 to be small in the kinematic region accessible to experiment. Even at
 ultra high energies, shown in \fig{h}-b, $  {\cal H}\Lb Y,y_1,y_2\Rb\,
\,\leq\,\,0.012$.   
 
 On the other hand, the contribution of \eq{REL12} is small and is
 proportional to $G_{3\pom}/g_i\,\ll\,1$. Bearing in mind that $G_{3
 \pom} = 1.29 $ in our approach, one can see that maximum value for 
 $ \mbox{max}\Lb {\cal H}\Rb \approx 0.012$ and the values of 
 \beq \label{VALSP}
  \mbox{max}\Lb {\cal H}\Rb \frac{G_{3 \pom}}{g_1\Lb 0
 \Rb}\,\approx \,1.4 \,10^{-4};~~~~~~~~~~\mbox{max}\Lb
 {\cal H}\Rb \frac{G_{3 \pom}}{g_2\Lb 0 \Rb}\,\approx 1.4 \,10^{-3}; \eeq
 
 Therefore, we expect that the contribution of the correlations due 
 semi-enhanced diagrams, is negligibly small.

\begin{figure}[ht]
    \centering
  \leavevmode
      \includegraphics[width=15cm]{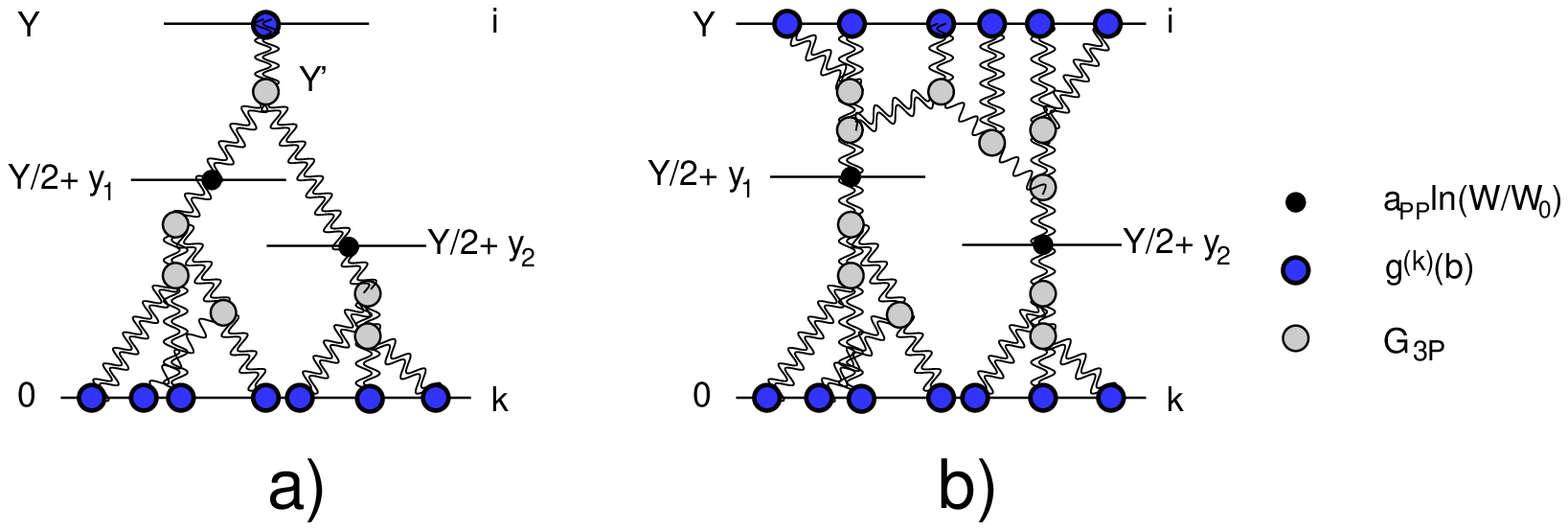}  
      \caption{ Mueller diagrams for the  rapidity correlation between
 two particles produced in one parton showers:   semi-enhanced diagrams.
  \fig{nimcor1shsen}-a  shows the first Mueller diagram, while
 \fig{nimcor1shsen}-b gives the structure of general diagrams.
  The double  wavy lines describe   dressed BFKL Pomerons.
 The blobs stand for the vertices  shown in the legend}
\label{nimcor1shsen}
   \end{figure}

 The general expression for the double inclusive cross section
 (see \fig{nimcor1shsen}-b) can be written using 
  two new functions $S^i\Lb y_1,y_2\Rb$  and $S_k\Lb y_1, y_2\Rb$
  defined as
 \bea 
&& S^i\Lb y_1,y_2\Rb\,\,=\label{1SHSEN1}\\
 &&\int d^2 b' \,\,N^{BK}\Bigg( g^{(i)}\,S\Lb m_i,b'\Rb
 \tilde{G}^{\mbox{\tiny dressed}}\Lb\h Y + y_1 \Rb\Bigg)\,\,N^{BK}
\Bigg( g^{(i)}\,S\Lb m_i, b'\Rb \tilde{G}^{\mbox{\tiny dressed}}\Lb\h
 Y + y_2 \Rb\Bigg) \times \Bigg( \frac{G_{3 \pom}}{ g^{(i)}
 \,S\Lb m_i,b'\Rb}\Bigg)\nn \\
 && S_i\Lb y_1,y_2\Rb\,\,= \label{1SHSEN2}\\
 &&\,\,\int d^2 b' \,\,N^{BK}\Bigg( g^{(i)}\,S\Lb m_i,b'\Rb
 \tilde{G}^{\mbox{\tiny dressed}}\Lb\h Y- y_1 \Rb\Bigg)\,\,N^{BK}\Bigg( g^{(i)}\,S\Lb m_i, b'\Rb \tilde{G}^{\mbox{\tiny dressed}}\Lb\h Y - y_2 \Rb\Bigg) \nn
  \eea
  
 It takes the form
  \bea \label{1SHSEN3}
&&\frac{d^2 \sigma^{\mbox{\tiny 1 parton shower}}_{\mbox{\tiny
 semi-enhanced}}}{ d y_1\,\,d y_2}\,\,=\,\, {\cal H}\Lb Y,y_1,y_2\Rb\Lb
 a_{\pom \pom}  \,\ln\Lb W/W_0\Rb\Rb^2\,\,\Bigg\{2\, \alpha^4 \,S^1\Lb
 y_1, y_2 \Rb\,S_1\Lb y_1, y_2 \Rb \nn\\
&& + \alpha^2\,\beta^2\,\Lb  S^1\Lb y_1, y_2 \Rb\,S_2\Lb y_1 
 y_2 \Rb\,\,+\,\, S_1\Lb -y_1,-y_2\Rb\,S^2\Lb - y_1,-  y_2 \Rb\Rb\,\,+\,\,
2\,\beta^4\,S^2\Lb y_1, y_2 \Rb\,S_2\Lb y_1, y_2 \Rb\Bigg\}
 \eea

  \subsection{Correlations in one parton shower: enhanced diagrams}
\begin{figure}[ht]
    \centering
  \leavevmode
      \includegraphics[width=15cm]{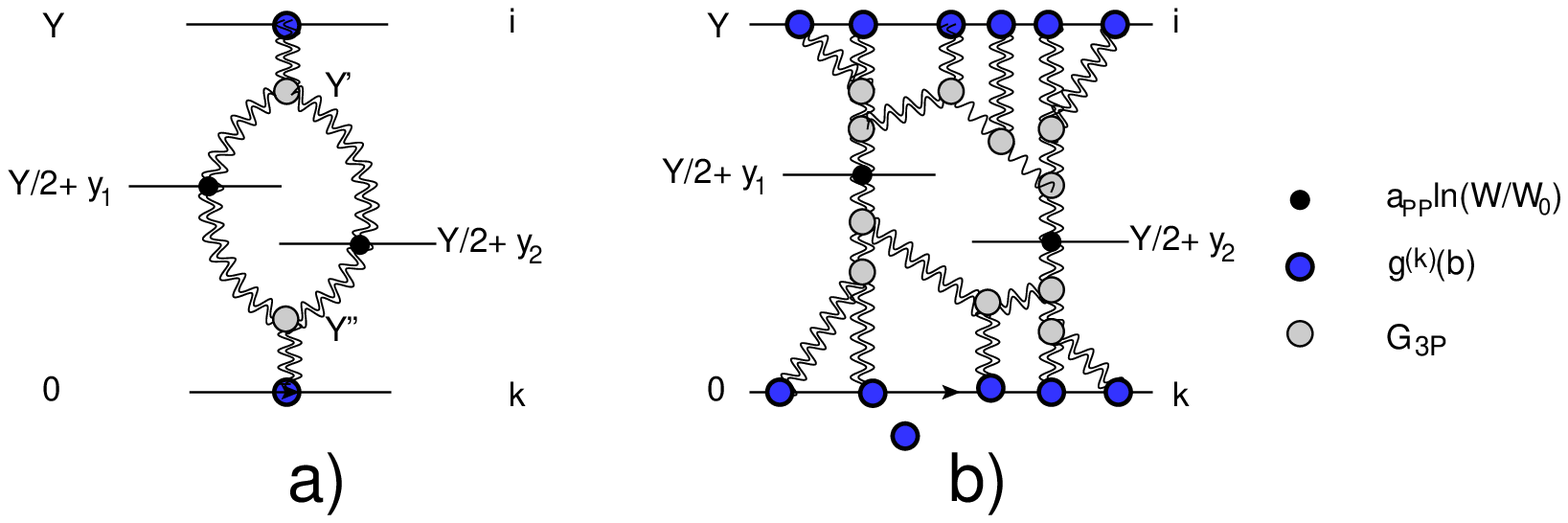}  
      \caption{ Mueller diagrams for the  rapidity correlation between
 two particles produced in one parton showers:  enhanced diagrams.
 \fig{nimcor1shen}-a  shows the first Mueller diagram, while
 \fig{nimcor1shen}-b indicates the structure of general diagrams.
  The double wavy lines describe dressed  BFKL Pomerons.
 The blobs stand for the vertices which are shown in the legend}
\label{nimcor1shen}
   \end{figure}

  The first Mueller diagram for the correlations from the enhanced
 diagram is shown in \fig {nimcor1shen}  -a, and has the following form
  
  \bea \label{1MDen}
 &&\frac{d^2 \sigma^{\mbox{one parton shower}}_{enh}}{d
 y_1\,d y_2}\,\,=\,\,\Lb a_{\pom \pom}\,\ln\Lb W/W_0\Rb\Rb^2
\sum_{i=1,k=1}^2\alpha^2_i\,\alpha^2_k\,\int^Y_{\max(\h Y - y_1,
 \h Y - y_2)}\!\!\!\!\! d Y'\,\int^{\min(\h Y - y_1, \h Y -
 y_2)}_0\!\!\!\!\! d Y'' \\
 &&\,\int d^2 b \,g_i\Lb b \Rb
 \tilde{G}\Lb Y - Y'\Rb\,G_{3 \pom}\,\tilde{G}\Lb Y' - \h Y
 - y_1\Rb \,\tilde{G}\Lb Y' - \h Y - y_2\Rb \nn \\
 &&,\int d^2 b' \,g_k\Lb b' \Rb
 \tilde{G}\Lb \h Y + y_2 -Y''\Rb\,\tilde{G}\Lb  \h Y + y_1
 - Y''\Rb \,G_{3 \pom}\,\tilde{G}\Lb Y''\Rb \nn \\
\eea  
  this can be re-written  as
  
  \bea \label{REL11}
 \frac{d^2 \sigma^{\mbox{one parton shower}}_{enh}}{d
 y_1\,d y_2}\,\,&=&\,\,\int d^2 b \,\frac{G_{3\pom}}
 {g_i\Lb b\Rb} \frac{G_{3 \pom}} {g_k\Lb b\Rb} \frac{d^2
 \sigma^{\mbox{2 parton showers}}}{d y_1\,d y_2\,d^2 b}    \eea  
  
  An example of  typical diagrams is shown in \fig{nimcor1shen}.
  The formula, summing all diagrams shown in \fig{nimcor1shen}-b takes the form
    \bea \label{1SHEN1}
&&\frac{d^2 \sigma^{\mbox{\tiny 1 parton shower}}_{\mbox{\tiny enhanced}}}{
 d y_1\,\,d y_2}\,\,= \,\,{\cal K}\, \Lb a_{\pom \pom} 
 \,\ln\Lb W/W_0\Rb\Rb^2\, {\cal H}\Lb Y,y_1,y_2\Rb\,{\cal
 H}\Lb Y,-y_1,-y_2\Rb\,\,\\&&\Bigg\{\alpha^4 \,S^1\Lb y_1,
 y_2 \Rb\,S^1\Lb -y_1, -y_2 \Rb \,\, + \alpha^2\,\beta^2\,\Bigg[S^1\Lb
 y_1, y_2 \Rb\,S^2\Lb -y_1 ,- y_2 \Rb\,\,+\,\,  S^2\Lb y_1, y_2
 \Rb\,S^1\Lb -y_1 ,- y_2 \Rb\Bigg]\nn\\
&&+\,\,
\beta^4\,S^2\Lb y_1, y_2 \Rb\,S^2\Lb -y_1, -y_2 \Rb\Bigg\}
 \eea
   where 
   
   \beq \label{1SHEN2}
      {\cal K}\,\,=\,\,\int d^2 b  \Big(G^{\mbox{\tiny dressed}}\Lb
 Y,b\Rb\Big)^2\Bigg{/}\Big(\int d^2 b \,G^{\mbox{\tiny dressed}}\Lb
 Y,b\Rb\Big)^2   
   \eeq
   
  where $G^{\mbox{\tiny dressed}}\Lb Y,b\Rb  $ is determined by \eq{G}
 and \eq{T}. The contributions of enhanced diagrams are proportional to
 the square of the ratios given by \eq{VALSP} and, therefore, they are
 negligibly small.
  
 \subsection{Correlation function}
 
 The correlation function $R\Lb y_1,y_2\Rb$ is defined as
 \beq \label{R1}
 R\Lb y_1,y_2\Rb\,\,=\,\,\sigma_{NSD}\,\Bigg\{  \frac{d^2 \sigma^{\mbox{\tiny 2 parton showers}}}{ d y_1\,\,d y_2}\,\,+\,\,\frac{d^2 \sigma^{\mbox{\tiny 1 parton shower}}_{\mbox{\tiny semi-enhanced}}}{ d y_1\,\,d y_2}\,\,+\,\, \frac{d^2 \sigma^{\mbox{\tiny 1 parton shower}}_{\mbox{\tiny enhanced}}}{ d y_1\,\,d y_2}\Bigg\}\Bigg{/}
 \Bigg\{ \frac{d \sigma}
 {d y_1}\,\frac{d \sigma}
 {d y_2}\Bigg\}\,\,\,-\,\,\,1
 \eeq
 \subsection{Kinematic corrections}
\begin{figure}[ht]
    \centering
  \leavevmode
      \includegraphics[width=8cm]{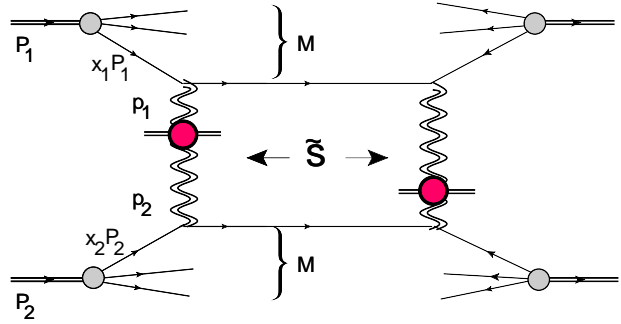}
      \caption{ The Mueller diagram for the  rapidity correlation between
 two particles produced in two  parton showers: the diffractive cut of
 the diagram.
  The double wavy lines describe dressed  BFKL Pomerons. The blob stands
 for $a_{\pom \pom}$.}
\label{kin}
   \end{figure}

\begin{figure}[ht]
    \centering
  \leavevmode
      \includegraphics[width= 11.3cm]{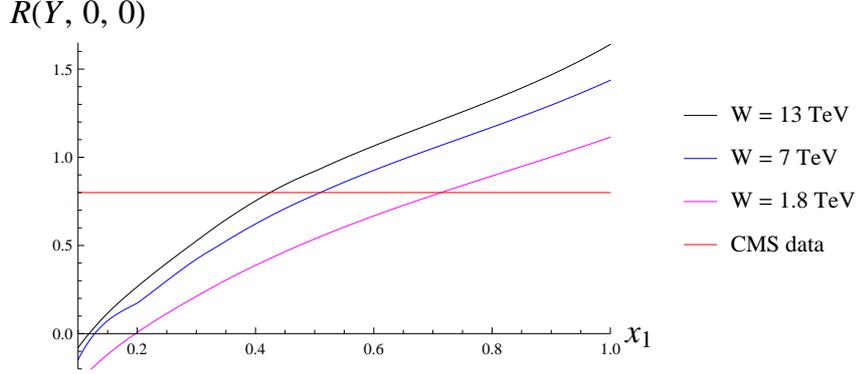}
      \caption{ $R\Lb Y,0,0\Rb$ versus $x_1$ for $Y =\ln\Lb 
x^2_1W^2/W^2_0\Rb$
 for different energies $W$. The red line indicates the moment
 $C_2=\langle|n^2|\rangle/\Lb\langle|n^2|\rangle\Rb^2$ = 2 for   
 the window in rapidity $-0.5 <y < + 0.5$, measured by the CMS
 experiment \protect\cite{CMSMULT}.}
\label{m}
   \end{figure}

In  all our previous equations we  assumed that $Y =
 \ln\Lb W^2/W^2_0\Rb$ with $W_0 = 1\,GeV$. This assumption appears
 natural for the elastic amplitude, and the cross section of the
 single inclusive production, but it should  be re-examined for the
 correlation function. For this observable, the  definition of $Y$
  has to be modified to account for the fact,
that the energy of the parton shower is not equal to  $W = \sqrt{s}$
 ($s = W^2$),
but it is smaller or equal to $\tilde{s}\,=\, \tilde{W}^2\,=x_1 x_2
 s\,=\,x_1 x_2 W^2$
(see \fig{kin}, where we show the diffractive cut of the Mueller
 diagram of \fig{nimcor2sh}-a).
 The simplest way to find $x_1$ and $x_2$ is to assume
that both $p^2_1 = p^2_2 = - \bar{Q}^2  \gg  \mu^2_{\mbox{soft}}$, where
$\mu_{\mbox{soft}}$ is the scale of the soft interactions,
$\mu_{\mbox{soft}}\, \sim \,\Lambda_{QCD}$. In our approach the scale 
 of hardness for the BFKL Pomeron $\bar{Q} = m \gg \,\mu_{\mbox{soft}}$. 
Bearing this in mind,
the energy variable $x_1$ ($x_2$) for gluon-hadron scattering
is equal to
\bea \label{X} 
&&0  \,=\, (x_1\,P_1 + p_1)^2\, =\, -\, \bar{Q}^2 \,
+ \,x _1\,2\, p_1\cdot P_1; ~~~~~~~   p^2_1\,
=\, -\, \bar{Q}^2; ~~~~~~~~~~~~~~x_1\,
=\,\frac{\bar{Q}^2}{ M^2 + \bar{Q}^2}.
\eea
$p_1$, $P_1$ and $x_1P_1$ are the momenta of the
gluon, the hadron and the parton (quark or gluon) with which the
 initial gluon interacts. From \eq{X} one  has that
\beq \label{STILDE}
\tilde{s}\,\,=\,\,x_1 x_2 S \,\,=\,\,\frac{s \,\bar{Q}^4}{M^4}
\eeq

where $M$ denotes the mass of produced hadron in the diffractive process.
 For the two channel model, it is the mass of the diffractive state.
 We can use the quark structure function to estimate the typical value
  of $x_1 = x_2$ as it is suggested in Ref.\cite{GLMDM}. Using the
 structure functions at $Q^2 \approx 25\,GeV^2$ one finds that
 $\langle|x_1|\rangle \approx 0.3 \div 0.5$. In \fig{m} the values
 of $R\Lb Y,0,0\Rb$ are plotted for $Y = \ln\Lb x^2_1W^2/W^2_0\Rb$
 as function of $x_1$.

 \subsection{Correlation in  one parton shower: emission from one BFKL Pomeron}
\begin{figure}[ht]
    \centering
  \leavevmode
      \includegraphics[width=9cm]{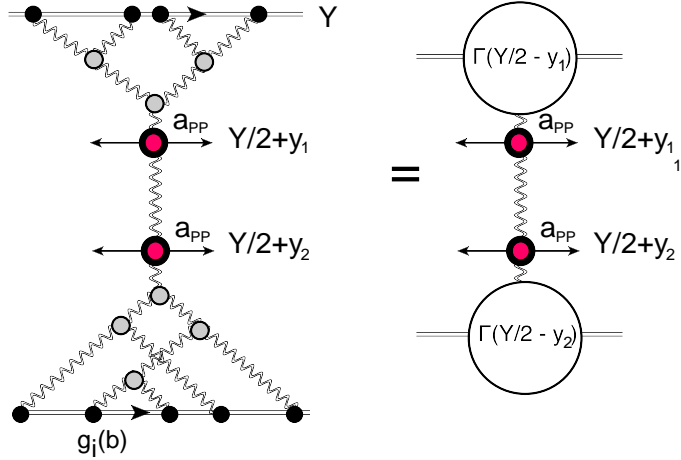}
      \caption{ Mueller diagram for emission of particles from
 one BFKL Pomeron.}
\label{pom}
   \end{figure}

  In addition to the sources of correlation that have been discussed above,
 we need to take into account  the correlation between two gluons 
emitted from one
 BFKL Pomeron (see \fig{pom}). At large $ y_{12} = |y_1 - y_2|$ the diagram
 of \fig{pom} induces  long range correlations in 
  rapidity, however, at 
small
 $y_{12}$ this emission is suppressed, and we do not expect a 
large contribution
 from this source.
  
  The contribution of this diagram can be written in the form
  \bea \label{POM1}
&&  R^{\mbox{\tiny BFKL}}\Lb y_1,y_2\Rb\,\,=\\
&&\,\,\sigma_{NSD}\frac{\sum^2_{i,k =1}\,\alpha_i \,\alpha_k \Gamma_i\Lb \h Y - y_1\Rb \tilde{G}^{\mbox{\tiny dressed}}\Lb y_{12}\Rb
  \Gamma_k\Lb \h Y + y_2\Rb  }{\sum^2_{i,k =1}\,\alpha_i \,\alpha_k\Gamma_i\Lb \h Y - y_1\Rb\,  \Gamma_k\Lb \h Y + y_1\Rb \,\sum^2_{i,k =1}\,\alpha_i \,\alpha_k \Gamma_i\Lb \h Y - y_2\Rb\,  \Gamma_k\Lb \h Y + y_2\Rb   }\nn
  \eea

  where 
  \beq \label{POM2}
   \Gamma_{i}\Lb  y \Rb  \,=\,\int d^2 b \,N^{BK}\Lb g_i\Lb b \Rb 
  \tilde{G}^{\mbox{\tiny dressed}}\Lb y\Rb  \Rb
   \eeq
  
   For small values of the argument
 $  \Gamma_{i}\Lb  y \Rb \,   \xrightarrow{y \ll 1}\,\int d^2 b
 \, g_i\Lb b \Rb   \tilde{G}^{\mbox{\tiny dressed}}\Lb y\Rb   $
 \eq{POM1}, has no dependence on $y_{12}$, leading to long range
 rapidity correlations. However, it turns out that the exact computation,
 leads to very small values of $  R^{\mbox{\tiny BFKL}}\Lb y_1,y_2\Rb$:
 approximately 0.2 $\div$ 0.4 \% of the contributions from the sources
 discussed above.

 \subsection{Short range rapidity correlation}
\begin{figure}[ht]
    \centering
  \leavevmode
      \includegraphics[width=9cm]{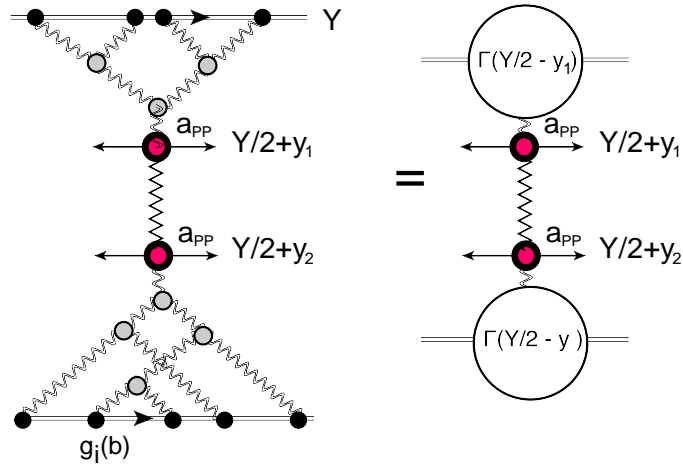}
      \caption{ Mueller diagram for the short range rapidity correlations.
 Wavy double lines denote BFKL Pomerons,  zigzag lines describe the 
exchange of the
  secondary Reggeon trajectory.}
\label{pomreg}
   \end{figure}

Besides  long range rapidity correlations, the emission
 from one BFKL Pomeron, as well as the hadronization in one
 gluon jet, can lead to  short range correlations in
 rapidity. Unfortunately, at present, this contribution cannot be
 treated on pure theoretical grounds, as
 it is involves,  confinement effects. To estimate this contribution, we 
introduce
the  Mueller diagram, shown in 
\fig{pomreg}, where we
  describe this correlation by the phenomenological
 constant $a_{\pom \reg}$, and introduce the correlation length
 $\Delta \approx 2$. In the diagram of \fig{pomreg} for  the zigzag
 line we have $a^2_{\pom \reg}\exp\Lb - \frac{y_{12}}{\Delta}\Rb$ ($y_{12} = | y_1 = y_2|$).
 Our estimate for $\Delta$ stems from Reggeon phenomenology, in which
 the zigzag line describes the contribution of the secondary Reggeon,
 with a propagator $\exp\Lb - ( 1 - \alpha_\reg(0)) y_{12}\Rb$ and
 $\alpha_\reg(0) \,\approx \, 0.5$.
  
The contribution of this diagram  takes the form
  \bea \label{REG1}
&&  R^{\mbox{\tiny short range}}\Lb y_1,y_2\Rb\,\,=\,\,\\
&&\Lb \frac{a_{\pom \reg}}{a_{\pom \pom}}\Rb^2\,\sigma_{NSD}
\frac{\sum^2_{i,k=1} \alpha_i \,\alpha_k\,\Gamma_i\Lb \h Y -
 y_1\Rb e^{- \frac{y_{12}}{\Delta}}\,
  \Gamma_k\Lb \h Y + y_2\Rb  }{\sum^2_{i,k=1} \alpha_i
 \,\alpha_k\,\Gamma_i\Lb \h Y - y_1\Rb\,  \Gamma_k\Lb
 \h Y + y_1\Rb \, \sum^2_{i,k=1} \alpha_i \,\alpha_k\,\Gamma_i\Lb
 \h Y - y_2\Rb\,  \Gamma_k\Lb \h Y + y_2\Rb }\nn
  \eea

 \section{Predictions and comparison with the experiment}
 \fig{m}, shows that the correlation function increases with
 energy, and  becomes rather large (of the order of 1) at $W = 7\,TeV$. 
This
 qualitative feature is in agreement with the experimental data from the 
LHC.
 The first set of data, is the multiplicity distribution measured by the 
CMS
 collaboration\cite{CMSMULT}. In particular,  the value of
 $C_2=\langle|n^2|\rangle/\Lb\langle|n^2|\rangle\Rb^2$  turns   
 out to be very close to 2, for the window in rapidities $-0.5 <
 \eta < 0.5$. Since $C_2 = R\Lb 0,0\Rb + 1 + 1/dN/d\eta|_{\eta=0}$
 where $d N /d \eta_{\eta=0}$ denotes  the multiplicity at $\eta=0$, and
 at $W = 7\, TeV$  it is equal to 5.8, 
while
  $R\Lb 0, 0\Rb = 0.82$\footnote{In this section we use pseudo  
 rapidity $\eta$ instead of rapidity $y$, since this variable is used
to display data from LHC
  experiments.  We recalculate $\eta = h\Lb y\Rb$ where function
 $h$ is taken from our paper \cite{GLMINCL}.}.
 
 The second set of the data is the measurement of the double parton
 interaction (DPI)\cite{DPI}. In the LHC experiments, the double inclusive
 cross sections of two pairs of back-to-back jets with momenta $p_{T,1}$
 and $p_{T,2}$,  were measured with rapidities  of two pairs ($y_1$ and
 $y_2$)  close to each other ($y_1 \approx y_2$). These pairs
 can only be produced  from two different parton showers.  The data 
were 
 parameterized in the form
\beq \label{XSEFF}
\frac{d \sigma}{d y_1 d^2 p_{T,1} d y_2 d^2 p_{T,2}} \,\,=\,\,\frac{m}{2
 \sigma_{eff}}\,\frac{d \sigma}{d y_1 d^2 p_{T,1}}\frac{d \sigma}{d y_2
 d^2 p_{T,2}}
\eeq
where $m =2 $ for pairs of different  jets, and $m=1$ for identical 
pairs.
  One can calculate the rapidity correlation function using \eq{XSEFF} 
\beq \label{RP}
R\Lb y_1, 
y_2,p_{T,1},p_{T,2}\Rb\,\,=\,\,\frac{\frac{1}{\sigma_{in}}\frac{d
 \sigma}{d y_1 d^2 p_{T,1} d y_2 d^2 
p_{T,2}}}{\frac{1}{\sigma_{in}}\frac{d
 \sigma}{d y_1 d^2 p_{T,1}}\,\frac{1}{\sigma_{in}}\frac{d \sigma}{d y_2 
d^2
 p_{T,2}} }\,\,-\,\,1\,=\frac{\sigma_{in}}{\sigma_{eff}}\,\,-\,\,1\,\,  
\approx\,\,2
\eeq
For the above the estimates we use $\sigma_{eff} $=\,12 - 15\,mb (see 
Refs.
 \cite{DPI}) and $\sigma_{in} = \sigma_{tot} - \sigma_{el} - \sigma_{sd} -
 \sigma_{dd} \,\approx $\,50 mb for the energy $W = 7 \,TeV$
 (see Ref.\cite{GLM2CH} and references therein).   These data
 confirm that at high energies we are dealing   with a 
system of partons
 that have a large mutual attraction . The fact that we predict 
 a smaller    
 correlation than in this experiment, does not discourage us, since the
 correlation function in \eq{RP}  differs from the one that we
 calculate (see \eq{R}).

The forward-backward correlation has been measured by the ATLAS
 collaboration in Ref.\cite{ATLASCOR}. The observable that was
 used in Ref.\cite{ATLASCOR} , differs from the correlation function
 $R\Lb \eta_1,\eta_2\Rb$, and it can be re-written as
 
\beq \label{FBW}
\rho^n_{f b}\,\,=\,\,\frac{R\Lb \eta_1,\eta_2\Rb}{\sqrt{R\Lb
 \eta_1,\eta_1\Rb\,R\Lb \eta_2,\eta_2\Rb}}
\eeq
 The value  of $\rho^n_{f b}\,\sim\,0.666$ \cite{ATLASCOR}
  indicates large correlations, but it is difficult to compare
 $\rho^n_{f b}$ with our estimates, since ATLAS introduced a specific
  selection: the $p_T$ of all produced particles should be larger than 
500\,MeV, 
while
 $R\Lb \eta_1,\eta_2\Rb$ is defined as integrated over all momenta.

\begin{figure}[ht]
    \centering
  \leavevmode
      \includegraphics[width=10cm]{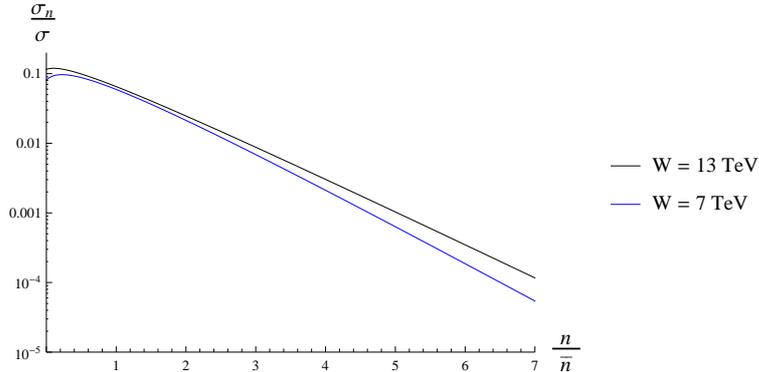}
      \caption{ $\sigma_n/\sigma$ versus  multiplicity $n/\bar n$ at  
 different energies $W$ in the rapidity window $-0.5 < y < 0.5$.}
\label{n}
   \end{figure}

 Using $R\Lb 0,0\Rb$, we estimate the values for $C_2$ for $W=13\,TeV$,
 and using the formula for the negative binomial distribution:
 \beq \label{SN}
 \frac{\sigma_n}{\sigma}\,=\,\Lb \frac{r}{\bar  n + r}\Rb^r  
 \frac{\Gamma\Lb n + r\Rb}{n!\,\Gamma\Lb r \Rb}\Lb \frac{\bar
 n}{\bar n + r}\Rb^n
 \eeq
 we obtain the multiplicity distribution in the rapidity window $-0.5 <
 \eta < 0.5$ shown in \fig{n}. In \eq{SN} $\bar n = d N/d \eta|_{\eta = 
0}$
 which was calculated in our paper \cite{GLMINCL}. From \fig{n}  
  we expect a violation of the KNO scaling behavior\cite{KNO}.
Accordingly to  KNO scaling $\sigma_n/\sigma = F\Lb n/\bar n\Rb$ with
 $\bar n  = \int^{0.5}_{-0.5}\, d \eta\, dN/d\eta = dN/d \eta|_{\eta=0}$.

It turns out that $R\Lb y_1,y_2\Rb$ at fixed energy, depends 
 neither on $y_1$ nor on $y_2$, giving a perfect example of long
 range  correlations in rapidty.  To understand why we have 
these
 features, it is instructive to start from \fig{nimcor2sh}-a at small
 values of $Y$.  In this kinematic region we can replace
 $\tilde{G}^{\mbox{\tiny dressed}} \,\to\,\tilde{G}^{\mbox{\tiny bare}}
 \,=\,\tilde{T}\Lb Y\Rb$ and $N^{BK}\Lb Y, b \Rb \,\to\,g_i\Lb b\Rb
 \,\tilde{G}^{\mbox{\tiny bare}}$. After simple algebra, 
 the correlation function is equal to
\beq \label{SIMPLE}
R\Lb y_1, y_2\Rb \,=\,\frac{1}{\alpha^4}\,
 \sigma_{NSD}\,\frac{\int d^2 b\Big(\int d^2 b'
  g_1\Lb \vec{b} \Rb\, g_1\Lb \vec{b} - \vec{b}'\Rb\Big)^2}{ \Big(\int d^2 b\,
  d^2 b'\,  g_1\Lb \vec{b} \Rb\, g_1\Lb \vec{b} -
 \vec{b}'\Rb\Big)^2}\,\,-\,\,1
\eeq
In \eq{SIMPLE} we use the fact that in our model $g_1\,\gg\,g_2$. 
\eq{SIMPLE}
 leads to a correlation function that does not depend on $y_1$ and $y_2$. 

On the other hand, at very large $Y$, $N^{BK}\Lb Y, b \Rb
 \,\to\,\Theta\Lb R\Lb Y\Rb - b\Rb$, where $\Theta\Lb b\Rb $ is a step 
function.
Plugging in this simple expression, we obtain
\bea \label{SIMPLE1}
R\Lb y_1, y_2\Rb \,&=&\, \sigma_{NSD}\,\frac{\int d^2 b\, 
 \Theta\Lb R\Lb \h Y - y_2\Rb - b\Rb\,\Theta\Lb R\Lb \h Y
 - y_1\Rb - b\Rb\,R^2\Lb \h Y - y_1\Rb\,R^2\Lb \h Y
 - y_2\Rb}{\,\,\pi^2\,R^4\Lb \h Y - y_1\Rb\,R^4\Lb \h Y - y_2\Rb}\,\,-\,\,1\nn\\
&\,\to\,&\,\,\,\frac{ \sigma_{NSD}}{\pi  R^2\Lb \h Y -
 y_1\Rb\Big{|}_{y_1 \,<\,y_2}  }\,\,-\,\,1\eea

$R\Lb Y\Rb$ in \eq{SIMPLE1} denotes a typical impact parameter
 at large $Y$, which is proportional to $Y$ \footnote{We trust
 that our use of the same notation for the correlation  function and
 typical $b$, will not  confuse the reader}. Recall,
 that at high energies, all components of the wave functions
 in the two channel model, give the same contribution. This
 is the reason why we do not have an extra factor which depend
 on $\alpha$ and $\beta$.

     \begin{figure}[ht]
    \centering
  \leavevmode
      \includegraphics[width=10cm]{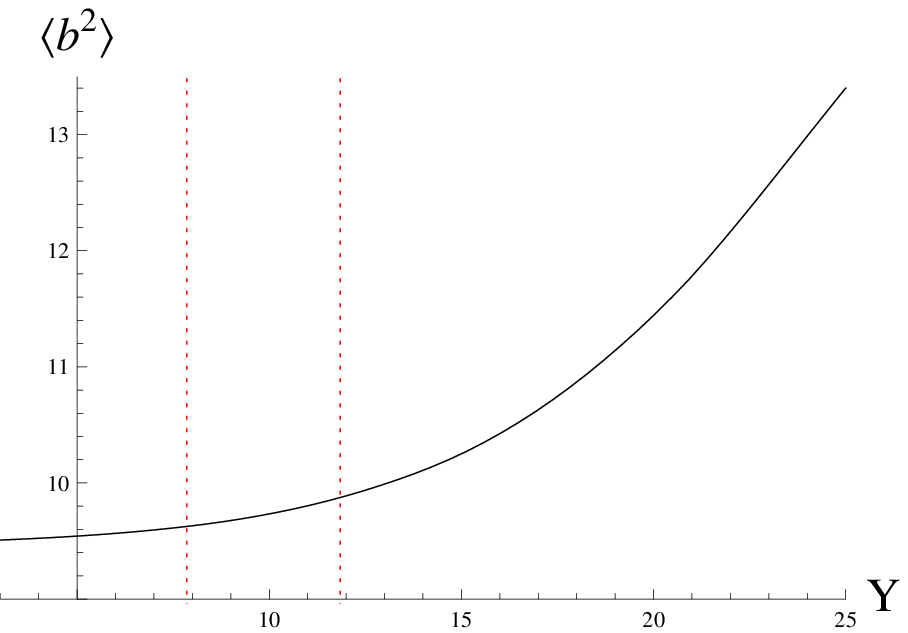}  
      \caption{ $\langle |b^2\Lb Y\Rb|\rangle$, defined in \protect\eq{B2},
 versus $Y$.}
\label{b2}
   \end{figure}

\eq{SIMPLE1} shows the logarithmic dependence on $y_1$.
 Using \eq{SIMPLE1} we can estimate the $y_1$($y_2$) dependence
 of $R\Lb y_1,y_2\Rb)$ calculating 
\beq \label{B2}
\langle |b^2\Lb Y\Rb|\rangle\,\,=\,\,\frac{\int b^2
 \,d^2 b \,N^{BK}\Lb g_i\Lb b\Rb \,\tilde{G}^{\mbox{\tiny dressed}}\Lb
 Y\Rb\Rb}{\int \,d^2 b \,N^{BK}\Lb g_i\Lb b\Rb \,\tilde{G}^{\mbox{\tiny dressed}}\Lb
 Y\Rb\Rb}
\eeq
From \fig{b2}, in which we plot the results of our calculation, one can
 see, that only at  large $Y \,>\,11$ does $ \langle |b^2\Lb Y\Rb|
\rangle$ start showing  visible $Y$ dependence. Two vertical dotted
 lines mark the widow in rapidity, which is  essential in  the calculation
 of the correlation function at $W = 7 \,TeV$ for $-2 \,<\,y_1,y_2
 \,<\,2$.  We can expect  a change of $R$   by 
2\%. The actual calculation gives even less.  Using $N^{BK}\Lb Y,
 b \Rb \,\to\,\Theta\Lb R\Lb Y\Rb - b\Rb$ gives $R^2\Lb Y\Rb =2
 \langle |b^2\Lb Y\Rb|\rangle$. Using this relation, we 
 estimate $R\Lb y_1,y_2\Rb$ as
\beq \label{SIMPLE2}
R\Lb y_1, y_2\Rb \,=\,\frac{1}{\alpha^4}\,\frac{ \sigma_{NSD}}{2 \pi
 \, \langle |b^2\Lb \h Y - y_1\Rb|\rangle} \,\,-\,\,1
\eeq
At $W = 7\,TeV$  from \eq{SIMPLE2} we find that $\,R\Lb y_1, 
y_2\Rb\,=\,1.647$, 
 while the exact calculation give 1.437 (see \fig{m}).
 At $W = 13 \,TeV$ this simple formula leads to $R\Lb
 y_1, y_2\Rb = 1.72$, versus 1.64 from the exact calculations (see 
\fig{m}).

The correlations,  measured by the ATLAS collaboration\cite{ATLASCOR},
 at first sight   contradict both  our estimate and the CMS
 data, regarding the multiplicity distribution.  We first  check 
\eq{FBW}.
 The measured observable has the form \cite{ATLASCOR}
\beq \label{nfb}
\rho^n_{f b}\,\,=\,\,\frac{\langle \Lb n_f - \langle n_f\rangle\Rb\,\Lb n_b
 - \langle n_b\rangle\Rb\rangle}{\sqrt{\langle \Lb n_f -
 \langle n_f\rangle\Rb^2\rangle\,\langle \Lb n_b- \langle
 n_b\rangle\Rb^2\rangle}}
\eeq
The numerator of \eq{nfb} can be written as $R\Lb y_1,y_2\Rb \Delta
 y_1\,\Delta y_2$, where $\Delta y_i$ is the interval of rapidities
 where  the hadrons are measured. However, at the same value of rapidity
 $R\Lb y_1,y_1\Rb( \Delta y_1)^2$ corresponds to $\langle n ( n - 1)\rangle -
 \langle n \rangle^2$. Therefore, the expression for
 $$\langle \Lb n_f - \langle n_f\rangle\Rb^2\rangle\,\,=\,\,\Lb R\Lb
 y_1,y_1\Rb +  \frac{1}{ \frac{d N}{d y_1} \,\Delta y_1}\Rb \Lb \Delta y_1\Rb^2
 $$
 which leads to the following formula for $\rho^n_{f b}$:
 \beq \label{nfbt}
 \rho^n_{f b}\,\,= \,\,\frac{R\Lb y_1, y_2\Rb}{\sqrt{ \Lb R\Lb y_1,y_1\Rb +
  \frac{1}{ \frac{d N}{d y_1} \,\Delta y_1}\Rb \, \Lb R\Lb y_2,y_2\Rb + 
 \frac{1}{ \frac{d N}{d y_2} \,\Delta y_2}\Rb }}
 \eeq 
 Taking $\Delta y_1 = 0.5$, we see that the first element of the Table 2 
is
 equal to 0.7, which is in good agreement with the experimental value 
$0.666
 \pm0.011$.
 
\begin{table}[ht]
{\small
\begin{center}
\begin{tabular}{|c|c|c|c|c|c|}
\hline
Forward $\eta$ interval\vstrut & 0.0 -- 0.5 & 0.5 -- 1.0 & 1.0 -- 1.5 & 1.5 --
 2.0 & 2.0 -- 2.5 \\
Backward $\eta$ interval & & & & & \\
\hline
0.0 -- 0.5 & 0.666 (0.70)& 0.624(0.643) & 0.592 (0.599) & 0.566 (0.565) & 0.540  (0.539)\\
           & $\pm0.011$ & $\pm0.011$ & $\pm0.011$ &  $\pm0.012$ & $\pm0.013$ \\
\hline
0.5 -- 1.0 & 0.624 (0.667) & 0.596(0.618) & 0.574 (0.580) & 0.553 (0.550) & 0.530 (0.527) \\
           & $\pm0.011$ &  $\pm0.011$ &  $\pm0.012$ &  $\pm0.013$ &$\pm0.014$ \\
\hline
1.0 -- 1.5 & 0.594 (0.640) & 0.576(0.596) & 0.560 (0.563) & 0.540 (0.537) & 0.518 (0.516) \\
           &  $\pm0.011$ &  $\pm0.012$ & $\pm0.013$ &$\pm0.014$ & $\pm0.014$ \\
\hline 
1.5 -- 2.0 & 0.571(0.615) & 0.557(0.577) & 0.544(0.548) & 0.526 (0.525) & 0.503 (0.508) \\
           & $\pm0.012$ &  $\pm0.013$ & $\pm0.014$ & $\pm0.014$ & $\pm0.016$ \\ 
\hline
2.0 -- 2.5 & 0.551 (0.593) & 0.540 (0.560) & 0.527 (0.535)& 0.507(0.515) & 0.487(0.499) \\
           & $\pm0.013$ & $\pm0.014$ &  $\pm0.014$ & $\pm0.016$ & $\pm0.018$ \\
\hline
\end{tabular}
\end{center}
}  
\caption{Multiplicity correlations for events at $\sqrt{s} = 7$~TeV
for events with a minimum of two charged particles in the kinematic
interval $p_{\rm T} > 100$~MeV and $|\eta| < 2.5$ for different
combinations of forward and backward pseudorapidity interval.  The data
 is take from Ref.\cite{ATLASCOR}.  The numbers in parentheses are  our 
estimates.  }
\label{tab:matrix}
\end{table}


   To describe our results given in Table 2,  
we need to also take  short range rapidity correlations into account.   In 
this table, 
in parenthesis,
 we have our estimates, which we obtain on describing the correlation 
function
 in the form

\bea \label{REXP}
&&R\Lb y_1,y_2\Rb\,\,=\,\,R^{\mbox{\tiny long range}}\Lb y_1,y_2\Rb\,\,+\,\,\Big(R^{\mbox{\tiny
 short range}}\Lb y_1,y_2\Rb\,-\,R^{\mbox{\tiny
 short range}}\Lb 0, 0\Rb\Big);\\
&& R^{\mbox{\tiny short range}}\Lb y_1,y_2\Rb\,=
 a\,\frac{\Gamma_1\Lb \h Y - y_1\Rb  e^{- \frac{y_{12}} {\Delta}}
  \Gamma_1\Lb \h Y + y_2\Rb  }{\Gamma_1\Lb \h Y - y_1\Rb\,  \Gamma_1\Lb \h Y + y_1\Rb \, \Gamma_1\Lb \h Y - y_2\Rb\,  \Gamma_1\Lb \h Y + y_2\Rb}  \nn
\eea
with $a =0.7$ and $\Delta = 2$. In \eq{REXP} $y_{12} = | y_1 - y_2|$ and we restrict ourself  by the contribution of the state "1"
in \eq{REG1}, since $g_1 \,\gg\,g_2$.
 We assumed that $  R^{\mbox{\tiny short range}}\Lb 0,0\Rb$=0 since at
 $y_{12}=0$ \eq{REG1}  leads to the long range correlations which we
 have calculated in section 3.4. One can see that the agreement is
 not perfect, but it demonstrates that the ATLAS data can be reproduced,
 by including the short range correlations.

 \section{Conclusion}
The main result of this paper is that in our model, which is based on the 
CGC/saturation approach,  we have discovered a 
mechanism that produces
 large, long range rapidity correlations at high energies.
  The large values of the correlation function $R\Lb y_1,y_2\Rb \,\geq \,1$
 at high energies, lends  strong support to the idea, that at high 
energies
 the system of partons that is produced, is not
  only dense, but also has  strong  attractive forces acting between the 
partons.
The resulting long range  rapidity correlations are
  independent  of $y_{1,2}$ . This prediction is in direct
 contradiction to the estimates from the soft  Pomeron based model 
that we  made (see Ref.\cite{GLMOLDCOR}). In that model the
 correlations from one parton  shower are larger than from the
 two parton showers, and they led to  the $y_{1,2}$ dependence. 
 Scrutinizing our formulae we found that the main reason for the
 smallness  of the correlation in one parton shower that we observed in
 our approach, stems from the most theoretically reliable part of our 
model:
 from the expression for the dressed BFKL Pomeron Green function. 

We demonstrated that our model is able to describe the LHC data,
 eminating from the CMS and ATLAS collaborations.  These data
   are certainly insufficient for  a thorough analysis of the details of
 our approach, but they confirm that the long range  rapidity 
correlations
 are large at high energies. The prediction for $W=13 \,TeV$ is shown in
 \fig{m}. The correlations should increase with the energy and the results
 measurements at $W = 13 \,TeV$ will clarify the situation.

In general we believe that this paper is the next natural  step,
 in building a model capable of  describing  soft high energy interactions 
based
 on the CGC/saturation approach.
            
  \section{Acknowledgements}
   We thank our colleagues at Tel Aviv university and UTFSM for
 encouraging discussions. Our special thanks go to  Carlos Cantreras , 
 Alex Kovner and Misha Lublinsky for elucidating discussions on the
 subject of this paper.
   This research was supported by the BSF grant   2012124 and  by  the
 Fondecyt (Chile) grant  1140842.

   \end{document}